\documentclass[conference]{IEEEtran}
\ifCLASSINFOpdf
\else
\fi
\usepackage[cmex10]{amsmath}
\usepackage{graphicx}
\begin{document}

\title{Stimulating Cooperation in Self-Organized Vehicular Networks}

\author{\IEEEauthorblockN{P. Caballero-Gil, J. Molina-Gil, C. Hern´andez-Goya and C. Caballero-Gil}
\IEEEauthorblockA{Department of Statistics, Operations Research and
Computing University of La Laguna\\
38271 La Laguna. Tenerife.Spain.\\
Email:\{pcaballe, jmmolina, mchgoya, candidoc\}@ull.es}}
\maketitle

\begin{abstract}
A Vehicular Ad-hoc NETwork (VANET) is a special form of Mobile Ad-hoc Network designed to provide communications among nearby vehicles and between vehicles and nearby fixed roadside equipment. Its main goal is to improve safety and comfort for passengers, but it can also be used for commercial applications. In this latter case, it  will be necessary to motivate drivers to cooperate and contribute to packet forwarding in Vehicle-to-Vehicle and Vehicle-to-Roadside communications. This paper examines the problem, analyzes the drawbacks of known schemes and proposes a new secure incentive scheme to stimulate cooperation in VANETs, taking into account factors such as time and distance. 

\end{abstract}

\section{Introduction}
\footnotetext{Work developed in the frame of the project TIN2008-02236/TSI supported by the Spanish Ministry of Science and Innovation and FEDER Funds.\\
Proceedings of APCC IEEE Asia Pacific Conference on Communications. Vol. 82 , (October 2009)	pp. 346-349.}


Vehicular ad hoc networks (VANETs) are important components of Intelligent Transportation Systems. The main benefit of VANET communication is seen in active safety systems that increase passenger safety by exchanging warning messages between vehicles. Other promising commercial applications are Added-Value Services such as: advertising support \cite{LPPGL07}, request/provide information about nearby companies, access to Internet, etc.

A VANET may be seen as a special type of ad-hoc network used to provide communications between On-Board Units (OBUs) in nearby vehicles, and between OBUs in vehicles and Road-Side Units (RSUs), which are fixed equipment located on the road.  
In particular, this paper deals with the topic of Inter-Vehicle Communication when the systems in a VANET do not rely on RSUs, and consequently constitute a Mobile Ad-hoc Network (MANET). 

The main advantage of VANETs is that they do not need an expensive infrastructure. However, their major drawback is the comparatively complex networking management system and security protocols that are required. This difficulty is mainly due to some specific characteristics of VANETs that allow differentiating them from the rest of MANETs such as their hybrid architecture, high mobility, dynamic topology, scalability problems, and intermittent and unpredictable communications. Consequently, these features have to be taken into account when designing any management service or security protocol. 

In order to bring VANETs to their full potential, appropriate schemes to stimulate cooperation need to be developed according to the specific properties and potential applications of VANETs. 
Many incentive schemes to stimulate cooperation in ad-hoc networks may be found in the bibliography \cite{HHHH08} \cite{LZ03} \cite{SA06} \cite{SNR03}. Some authors have made first approaches to the topic of cooperation in VANETs \cite{DFM05} \cite{FF06} \cite{WC07a} \cite{WC07b}. Related to the proposal here described, Buttyan and Hubaux proposed in \cite{BH03} and \cite{BH07}  the use of virtual credit in incentive schemes to stimulate packet forwarding. Also, Li et al. discussed some unique characteristics of the incentive schemes for VANETs in \cite{LW08} and proposed a receipt counting reward scheme that focuses on the incentive for spraying. However, the receipt counting scheme proposed there has a serious overspending problem. Based on the specific characteristics of VANETs, a more comprehensive weighted rewarding method is proposed here. 

In particular, the proposed scheme is based on incentives where the behavior of a node is rewarded depending on its level of involvement in the routing process. Schemes based on reputation were here discarded due to the high mobility of nodes in VANETs, which makes infeasible to maintain historical information about peers behavior.



Note that an important problem that must be dealt with in rewarding incentive schemes is the possibility for selfish or malicious users in the vehicles to exaggerate their contribution in order to get more rewards. In our proposal, we assign different possible incentives to vehicles according to their contribution in packet forwarding, in an effort to achieve fairness and provide stimulation for participation. Our scheme utilizes a weighted rewarding component to decide the specific incentive in each case so they help to keep the packet forwarding attractive to the potential intermediate vehicles. 

\section{Background}
A VANET may be seen as a variation of a MANET where the nodes are vehicles. In both types of networks, cooperation between nodes is required for the adequate performance, so there might be thought that  cooperation tools for MANETs can be also used for VANETs. In MANETs we can find two main approaches: Reputation-based schemes where packets are forwarded through the most reliable nodes, and Credit-based schemes where packet-forwarding  is dealt with as a service that can be evaluated and charged. In our work we have analyzed both schemes in order to find out whether they are suitable for VANETs.


\subsection{Why reputation based methods are not suitable?}
An important characteristic of VANETs is the high mobility of nodes. Taking into account such a parameter, it is impossible to establish a reputation-based scheme because it is infeasible to maintain historical information about peers' behavior in a VANET. This is because it is possible that two vehicles meet just once in a long period of time, and it is very difficult to listen whether a neighbor node actually forwards a packet.

\subsection{Why classical credit based schemes are not suitable?}
Incentive schemes have been proposed in order to solve cooperation problems in MANETs. The so-called {\it Packet Purse Model} where every source node puts a sum of money that it considers enough to reach the destination has an overspending problem because the source vehicle can not predict accurately the global size of the required reward.
Other known model is the so-called {\it Packet Trade Model} where the destination node pays the reward. In this case the model has problems because source nodes can send all the packets  they want as they have not to pay for them. This model produces a network overload. Consequently,  we can conclude that none of both schemes are a good solution for MANETs or for VANETs.

\section{Main Scheme}

\begin{figure}
 \centering
 \includegraphics[scale=0.6]{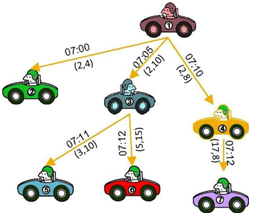}%
 \caption{Forwarding tree}
 \label{tree}
\end{figure}

Figure \ref{tree} shows a typical packet forwarding process in VANETs, here called {\it Forwarding Tree}. In such a figure several important features of routing in VANETs are represented: 
\begin{enumerate}
\item The root node corresponds to the source vehicle that first sprays the message.
\item Each intermediate vehicle corresponds to one node in the tree.
\item Each node ignores those packets that it had previously received. Consequently, every vehicle is present just once in every forwarding tree.
\item Each link in the tree corresponds to an encounter in the vehicular network, which is associated with a timestamp and the spatial coordinates indicating the position of the vehicles.
\end{enumerate}

According to the store-and-carry paradigm \cite{LW08} \cite{HCM09}, if an intermediate vehicle stores a packet for a long time or actively sprays the packet to other vehicles, the packet will be either more likely to reach the intended destination, or to arrive to more destinations, depending on the specific goal of the routing. Therefore, by simply combining storage time and number of sprays, we can define a useful contribution metric for the intermediate vehicles. In order to stimulate intermediate vehicles to contribute more, the source vehicle should reward each intermediate vehicle according to its contribution. 

Initially, the contribution $C_i$ to packet forwarding of a node $i$ during the forwarding process may be modeled as a linear convex combination balancing numbers of forwarding $f_i$ and the period the packet is stored $t_i$: 

$C_i=\alpha t_i + (1 - \alpha) f_i$.

However, this basic model implies a constant share reward $R$ which is promised for the source node to each intermediate node. This model may cause an overspending problem because the source vehicle cannot guess in advance the total reward since the number of nodes in the tree cannot be predicted easily. Such a problem might be solved maintaining constant the total reward and calculating the reward associated to each intermediate node $R_i$ after the packet reaches the destination according to the following formula: 

$R_i = \frac{R \cdot C_i }{C}$ where $C= \sum_i C_i$

When the packet reaches the destination, each node $i$ that participated in the forwarding should report its contribution $C_i$ to the source. The final contribution $C$ is calculated through the sum of the partial contribution of each node in the forwarding tree. Each intermediate node will receive $R_i$ as reward for forwarding. This model cannot be considered neither a good solution because selfish nodes might prefer keeping the packet rather than retransmitting it since they do not know in advance how much they can earn for forwarding and/or they might prefer not to share the reward. It happens when an intermediate node forwards the packet to a non final node because its proportional reward might decrease. 

In our first proposal we incorporate several parameters to be considered when dealing with rewarding. They are related to information such as packet delivery deadline and number of forwardings. In order to avoid that nodes prefer keeping the packet rather than retransmitting it, we consider that a packet should have a deadline depending on the characteristics of the information contained. If the sent information is added-value  information, then the deadline will be longer, whilst if the information is related to traffic safety, the deadline must be very short due to the urgency of transferring the information. The following notation is used to describe the parameters for the computation of rewards:
\begin{itemize}
\item Packet delivery deadline $T_{j}$.
\item Period $t_{ij}$ that packet $j$ is stored by  node $i$.
\item Number of forwardings $f_{ij}$ of packet $j$ by node $i$ before the deadline $T_{j}$.
\item Balancing factor $\alpha$
\end{itemize}

With these parameters we present our first proposal of contribution function:
$
C_{ij} =\alpha$($t_{ij}T_{j})+ (1-\alpha)$$f_{ij}$
where $\alpha \in (0,1).
$

In this first approach to the solution, the contribution of node $i$ for spreading packet $j$ is proportional to the time  $t_{ij}$ the packet is transported by the packet deadline $T_{j}$, and to the number of the forwardings $f_{ij}$. By counting the number of forwardings in this function, the objective  is to encourage nodes to forward packets and not to keep them without forwarding them. 

If the rate between time and deadline $t_{ij}/T_{j}$ were considered as factor, when the message is urgent and the deadline is short, the contribution would be higher. However in such a case, once the time $t_{ij}$   overpasses the deadline $T_{j}$, the  contribution still goes on increasing and even faster because the proportional factor is greater than 1. This effect can be corrected by using the inverse of the deadline so that the more urgent the message, the greater the value $1/T_{j}$. Therefore, when the time that the packet is stored passes the deadline the user contribution is no more increased. However, this is neither a good solution because although the deadline has been reached, the forwarding node continues getting a reward although it is a small amount.

Our second proposal tries to solve these problems. We propose a new contribution function in which three parameters are used, which can be interesting both for the source node and/or for the forwarding node. In particular we consider the following additional notation to describe the parameters for the computation of rewards: 
\begin{itemize}
\item Distance $d_{ij}$ between source and destination nodes when the packet $j$ is relayed by node $i$.  
\item Maximum distance $D_{j}$ where the information in the packet $j$ is considered interesting by the receivers. 
\end{itemize}

Each of the parameters considered in this convex function has a balancing factor, represented by $\alpha_1$, $\alpha_2$ and $\alpha_3$. The value that is assigned to each $\alpha_i$ depends on  the relevance that the source node prefers to assign to each component represented in the contribution function:

$
C_{ij} =\alpha_1$ $T_{j}(1-e^{-t_{ij}})+ \alpha_2$ $f_{ij}+ \alpha_3$ $(-D_{j}(1-e^{-d_{ij}})+D_j)$
where $\sum\limits_{k=1}^3 \alpha_k =1.
$

In the next subsections each part of this function is detailed, and both the justification why they are used and the repercussion they have in the contribution function are given.

\subsection{Time}
\label{time}
As discussed above, time is one of the most important parameters when trying to assure that a packet reaches the intended destination. If a vehicle stores a packet for a long time, it could  forward the package to more vehicles. However, this parameter could produce a selfish behavior because a node could prefer not to forward it and in this way not to share the final reward with potential forwarding nodes. This effect is avoided by considering in the proposed metric  the component associated to the following formula: 

$T_{j}(1-e^{-t_{ij}})$.

This function corresponds to the Stokes formula, which has a characteristic asymptotical behavior. This function is intended to set a maximum time $T_{j}$ that a node should store one packet. Note that the value of contribution increases when time increases. When $t_{ij}$ reaches the threshold $T_{j}$, the growth of contribution stops. In this way, a selfish behavior can be avoided because if the time threshold is properly set, those vehicles that retransmit the packet before the deadline will have increased their contribution.

\begin{figure}
  \centering
  \includegraphics[scale=0.35]{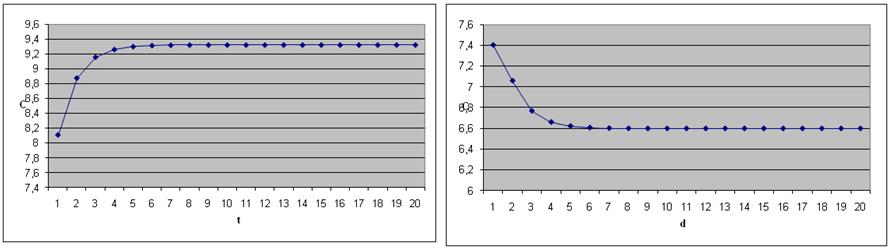}%
  \caption{Contribution versus time and distance}
  \label{cvt}
\end{figure}

Note in the example of Figure \ref{cvt} that the value of contribution increases when the time increases, and that when $t_{ij}$ reaches the threshold $T_{j}$ the contribution increase stops. In this way, both selfish behavior and forwarding after deadline are discouraged because vehicles that retransmit the packet before deadline will have their contribution increased.

\subsection{Forwarding}
The second term in the proposed contribution metric is related to the ultimate goal of our work. It deals with measuring the forwarding of packets by each intermediate node. This process is quite simple. It has not any restriction such as maximum or minimum possible values.	It consists of increasing the contribution of node $i$ to relay the packet $j$ by:

$f_{ij}$.


According to this factor, the more the vehicles collaborate in forwarding a packet, the bigger their final contribution is. In the proposed function, this parameter is the one that increases the contribution faster before the deadline. Consequently, the balancing factor $\alpha_2$ must be higher than the other two factors in order to encourage the forwarding of packets.

\subsection{Distance}
The evaluation of the effect of distance in the share rewarding process is the goal of the third term of the contribution function. This term has been incorporated thinking that in many cases information generated at a certain location  is not interesting out of a radius distance from than point. With this idea in mind, when the vehicles go too far from the  source of the original packet, this value decreases. 

For example, if we talk about an accident in a city center, it has not sense that the message reaches a neighbor city. Other possible situations where the same idea is applicable is where the information is sent by a commercial centre, hotel or restaurant, for instance. 


This term is similar to the one related to time commented in subsection \ref{time}. The goal is to obtain a function with asymptotical behavior that tends to zero when distance is near to $D_{j}$. The value $D_j$ is established by the source node. The expression that models this behavior is:

$-D_{j}(1-e^{-d_{ij}})+D_j$.


Figure \ref{cvt} shows an example where as the vehicle moves away from the source its contribution decreases, and when it reaches certain point it nulls. In this way, the vehicle does not get any benefit if it retransmits the packet outside the radio.

\section{Simulation Analysis}
\begin{figure}
  \centering
  \includegraphics[scale=0.40]{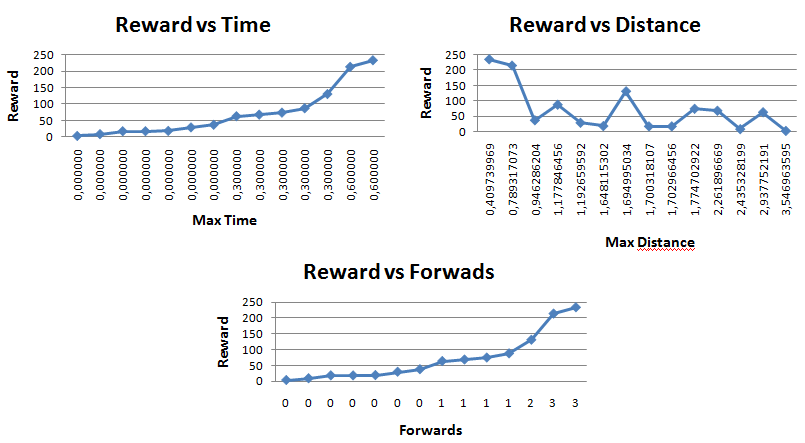}%
  \caption{Simulation}
  \label{Simulation}
\end{figure}

\begin{figure}
  \centering
  \includegraphics[scale=0.45]{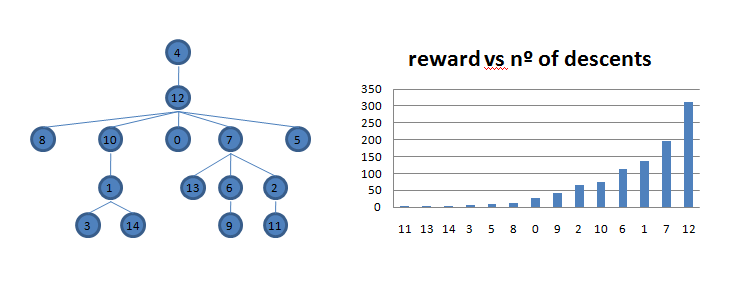}%
  \caption{Reward versus number of descents}
  \label{descents}
\end{figure}

In order to make a study of the proposal, several VANETs simulations have been implemented in NS-2. The NS-2 simulation parameters are the following: 15 nodes placed at random in an area of 800m x 800m. The range of action of each node is 100m. In each simulation, a node is randomly chosen and it starts sending a packet to its neigbours, who send it to all the nodes they meet inside their range of action. In Figure \ref {Simulation} we analyze the relationship among the rewards and the different parameters of the contribution function. According to the time plot, the scheme seems to send bigger rewards to those nodes who store packets for longer. However, note that these rewards are influenced by the number of forwards and the distance between the source node and the nodes forwarding the packet. In the forwards graph, the reward average increases according to the number of forwards. Finally, in the distance plot the scheme seems to give lower rewards to those nodes whose distance increases according to the source node initial position. For some nodes at a large distance, the reward average increased due to that their spray was bigger than the spray of the nodes in the same distance. This scheme provides more reward to the nodes that effectively sprayed the packet for a long time. Also according to Figure \ref {descents}, the proposal gives more reward to those nodes with higher contributions, which are usually those nodes that have more descendents. Consequently, cooperation among nodes is guaranteed thanks to the proposal. 

\section{Conclusion}
In this paper we have seen that a simple adaptation of known cooperation enforcement schemes defined originally for MANETs is not adequate to incentivize cooperation in VANETs. Consequently, we have proposed a new scheme where incentives are defined by a convex function that depends on different parameters. We have designed a metric for contribution according to the characteristics of VANETs and to parameters that are important both for source node and for enforcing cooperation among nodes. We conclude from our study that when designing these methods for distributing a reward, the parameters to be taken into account should be carefully assessed according to the network conditions.

Since this is a work in progress, many open questions exist such as the the analysis of how can data associated to traffic and weather conditions can be used in order to improve the efficiency of the proposal.

\section*{Acknowledgment}
Research supported by the Spanish
Ministry of Education and Science and the European FEDER Fund
under TIN2008-02236/TSI Project, and by the Agencia Canaria de Investigaci\'on, Innovaci\'on y Sociedad de la Informaci\'on under PI2007/005 Project.


\begin{thebibliography}{10}

\bibitem{BH03} Buttyan, L. and Hubaux, J.P.: Stimulating Cooperation in
Self-Organizing Mobile Ad Hoc Networks. ACM Mobile Networks and
Applications, 8(5), October (2003)

\bibitem{BH07} Buttyan, L. and Hubaux, J.P.: Security and Cooperation in Wireless
Networks. Cambridge Univ. Press (2007)

\bibitem{DFM05} Dotzer, F., Fischer, L. and Magiera, P.: VARS: A Vehicle Ad-Hoc Network Reputation System. Sixth IEEE International Symposium on a World of Wireless Mobile and Multimedia Networks. WoWMoM 2005. 13-16 June 2005 pp. 454-456 (2005) 

\bibitem{FF06} Fonseca, E. and Festag, A.: A Survey of Existing Approaches for Secure Ad Hoc Routing and Their Applicability to VANETS, Technical Report NLE-PR-2006-19, NEC Network Laboratories, March (2006)

\bibitem{HHHH08} Ho, Y.H., Ho, A.H.,  Hamza-Lup, G.L. and Hua, K.A.: Cooperation Enforcement in Vehicular Networks. International Conference on Communication Theory, Reliability, and Quality of Service, 2008. CTRQ'08. June 29-July 5 pp. 7-12 (2008) 

\bibitem{LPPGL07} Lee, S., Pan, G.,  Park, J.,  Gerla, M. and Lu, S.: Secure Incentives for Commercial Ad Dissemination in Vehicular Networks, MobiHoc'07, Canada, Sep 9-14 (2007)


\bibitem{LW08} Li, F. and Wu, J.: FRAME: An Innovative Incentive Scheme in Vehicular Networks. Proc. of IEEE International Conference on Communications (ICC) (2009)

\bibitem{HCM09} Hernandez-Goya, C., Caballero-Gil, P., Molina-Gil, J. and Caballero-Gil P.:Cooperation Enforcement Schemes in Vehicular Ad-Hoc Networks. Lecture Notes in Computer Science. EUROCAST. Vol: No. 5717, Spain Feb 15-20, (2009).

\bibitem{LZ03} Liu, P. and Zang, W.: Incentive-based modeling and inference of
attacker intent, objectives, and strategies, Proc. of the 10th
ACM Computer and Communications Security Conference (CCS'03),
Washington, DC, October pp. 179-189 (2003)

\bibitem{SA06} Shastry, N. and  Adve, R.S.: Stimulating cooperative diversity in
wireless ad hoc networks through pricing", IEEE Int. Conf. on
Communications, June (2006)

\bibitem{SNR03} Srinivasan, V.,    Nuggehalli, P. and Rao, R.R.: Cooperation in Ad Hoc
Networks, Proc. of Infocom, San Francisco, CA (2003)

\bibitem{WC07a} Wang, Z. and Chigan, C.: Countermeasure Uncooperative Behaviors with Dynamic Trust-Token in VANETs. IEEE International Conference on Communications, 2007. ICC'07. 24-28 June  pp. 3959-3964 (2007)

\bibitem{WC07b} Wang, Z. and Chigan, C.: Cooperation Enhancement for Message Transmission in VANETs. Wireless Personal Communications: An International Journal Volume 43,  Issue 1  October pp. 141-156 (2007)



\end{thebibliography}
\end{document}